# Extrapolative ML Models for Copolymers

Israrul H. Hashmi, Himanshu, Rahul Karmakar and Tarak K Patra*

Department of Chemical Engineering
Indian Institute of Technology Madras, Chennai 600036, India

**Abstract:**

Machine learning models have been progressively used for predicting materials' properties. These models can be built using pre-existing data and are useful for rapidly screening the physicochemical space of a material, which is astronomically large. However, ML models are inherently interpolative, and their efficacy for searching candidates outside a material's known range of property is unresolved. Moreover, the performance of an ML model is intricately connected to its learning strategy and the volume of training data. Here, we determine the relationship between the extrapolation ability of an ML model, the size and range of its training dataset, and its learning approach. We focus on a canonical problem of predicting the properties of a copolymer as a function of the sequence of its monomers. Tree search algorithms, which learn the similarity between polymer structures, are found to be inefficient for extrapolation. Conversely, the extrapolation capability of neural networks and XGBoost models, which attempt to learn the underlying functional correlation between the structure and property of polymers, show strong correlations with the volume and range of training data. These findings have important implications on ML-based new material development.

*Author to Correspond, E-mail: tpatra@iitm.ac.in

## I. Introduction

First principles-based computer simulations and experiments are limited by their inability to rapidly measure polymer properties, making them inadequate for efficiently screening the vast chemical and conformational space of polymers. However, recent progress in machine learning (ML) and the growing availability of data and software offer potential pathways to expedite polymer design.[1–5] Towards this end, significant progress has been made to create data-driven models that predict polymer properties. These models are built by using candidate polymers and labeling them by their properties, which are calculated using physics-based methods. A large variety of machine-readable fingerprints and chemical descriptors are proposed for such ML model development.[6–9] The fingerprint-property data are utilized for training and building ML models. Such ML models serve as a cheaper, albeit low-fidelity, surrogate for the high-fidelity first-principle-based simulations and experiments that are expensive. There exist a large number of numerical frameworks, such as random forest (RF), XGBoost and deep neural network (DNN) to build these ML models. While the potential of ML predictive models is very lucrative, there is a lack of clarity on how one should train an ML model to exhaustively fit a polymer's entire configurational and chemical space. As the mathematical framework of an ML model is flexible, the general perception is that it can be used to predict a wide range of properties, given that it is built upon a large number of data points. However, this hypothesis has not been rigorously tested.

We address the above issue of ML model development for a canonical polymer problem, viz. prediction of sequence-defined properties of a copolymer. The sequence of a copolymer appreciably controls its bulk and single-molecule properties, such as glass transition, ion transport, thermal conductivity, a single-molecule radius of gyration, and multimolecular aggregations.[10–16] The major challenge in constructing a universal model lies in a copolymer's overwhelmingly vast number of potential sequences. The sequence-specific nature is so profound that even a minor alteration in the copolymer sequence brings a substantial shift in many of its properties.[12,17–19] Often, the optimal properties are found in a sequence that appears non-intuitive and seemingly arbitrary, with an unknown level of sequence-specificity.[10,11,13] In recent times, MLs have been increasingly used to predict such sequence-defined properties of copolymers.[8,20–22] However, no agreed-upon strategy has emerged to decide the minimum sequence-property data required to build these models. Also, there is a lack of clarity on ML algorithms that are most suitable for the polymer sequence problem. Therefore, developing a model that captures a wide variability in sequence and property of a copolymer is challenging. Learning and predicting these variations of a copolymer required a large amount of data. As such, there is a lack of clarity on how much data is needed to accurately learn and predict sequence-properties correlations. Also, it appears that the volume of data required to build a high fidelity model is intricately connected to the learning algorithm. Perhaps, some algorithms need more data than others. Therefore, building a universal model remains a substantially complex task.

The objective of this work is to build ML models for extrapolation. To accomplish these objectives, we consider a representative problem, viz. copolymer adsorption free energy (AFE) on a surface. We chose three commonly used ML algorithms for predicting sequence defined property of polymers – viz., RF, XGBoost, and DNN. We build ML model using training data that represent a specific range of property and test their ability to predict the properties that are outside the training data. We consider two possible cases – i) the amount of training data increases for a specific region of the distribution, and ii) the range of training data increases starting from the central region of the distribution while the amount of data is fiexed. For both the cases, we analyze the performance of models for the out-of-distribution data. We show how the increment in the volume and range of training data of an ML model aid in generalizing its learning to unseen data, particularly the out-of-distribution data for a polymer problem. The extrapolation capability of an ML model is shown to vary with its learning algorithm. Although the current study focuses on sequence-property ML models of copolymers, we expect similar characteristics of ML models for other classes of materials and their properties. The current work has fundamental implications on the future development of ML models for materials property prediction and design.

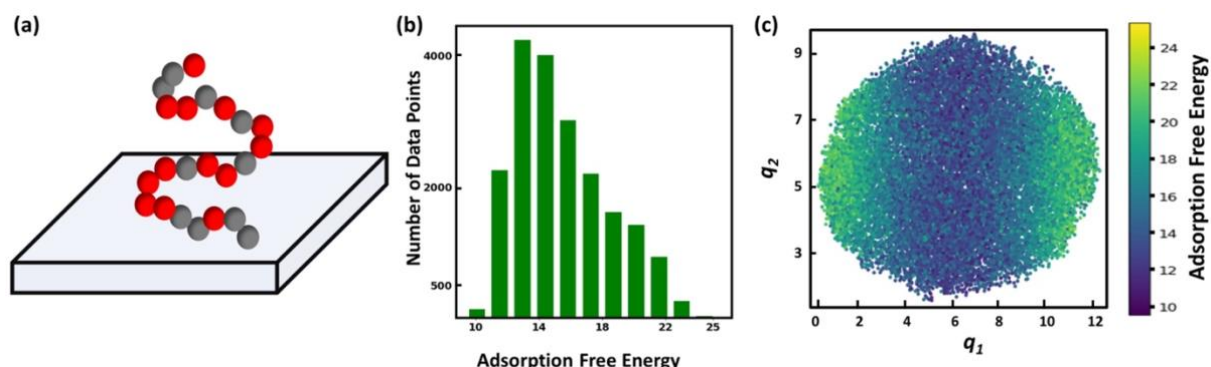

*Figure 1: Sequence-defined copolymer data. A schematic representation of a single copolymer chain adsorbed on a flat surface is shown in (a). The distribution of its adsorption free energy (AFE) are shown in (b). The projection of 20000 sequences on a 2D space via an UMAP is shown in (c).*

## II. Polymer Sequence-Property Data

The sequence-property data of a copolymer is taken from the work of Shi et al.,[23] wherein they reported the AFE of a copolymer on a functional surface made of chemically similar particles using coarse-grained molecular simulations. This simulation study has employed the Kremer-Grest bead-spring phenomenological model[24,25] to investigate the sequence-property correlation. In this model, two types of chemical moieties are linearly connected to form a copolymer. The interaction parameters of the moieties are adjusted to represent their chemical affinity in a given system. It is a well-established model for studying generic polymer properties in molecular simulations without considering specific polymer chemistry and conditions. This simple model is computationally very efficient and can be mapped to real polymers by tuning its parameters.[26] This data set consists of ~20000 sequences of a copolymer of chain length $N=20$ and their AFEs. A schematic representation of the model system is shown in Figure 1a, and the distribution of AFF of the dataset is shown in Figure 1b. The sequence of monomers in a polymer is represented as an array of binary numbers – 0 and 1 for the purpose of ML.[27] This binary array is also known as the feature vector or fingerprint for ML models. As the polymer chain is made of 20 monomers, the feature vector size is 20. A uniform manifold approximation and projection (UMAP) of all the binary fingerprints in the dataset are shown in Figure 1c. The UMAP[28] is a nonlinear dimensionality reduction technique that enables the visual inspection of high-dimensional sequence data of a copolymer in a 2D representation. The sequence and AFE distribution appear to be continuous in their respective spaces.

## III. ML Algorithms

All three ML models – RF, XGBoost and DNN are built within the Keras application programming interface (API).[29] The binary representation of a sequence of the copolymer chain serves as the input to the models, and the corresponding AFE of the polymer sequence on a surface is the output of the models. We define mean square error (MSE) as the loss function of an ML model, which represents the difference between the output of the model and the ground truth. Therefore, the loss function is written as $L(y, \hat{y}) = \frac{1}{N}\sum_{i=0}^{N}(y_i - \hat{y}_i)^2$. Here, $y$ and $\hat{y}$ are the actual and predicted AFE, respectively. Below, we briefly describe all three models and their development for predicting sequence-defined properties of a copolymer.

**Random Forest (RF).** An RF is an ensemble ML method that constructs decision trees.[30] The decision trees of a RF learn how to best split a dataset into smaller and smaller subsets to predict the target value. The splitting of a tree is based on the conditions that are applied on the input features. The conditions are represented as the leaves or nodes, and the possible outcomes are known as branches or edges. The splitting process continues until no further gain can be made or a preset rule is met, e.g. the maximum depth of a tree is reached. A schematic representation of a tree that is suitable for predicting polymer sequence-property data is shown in Figure 2. Here, the feature values are binary numbers – *0* and *1* that are arranged in a sequence. A randomly selected feature is used to split nodes and growing a tree. The growth of a tree (depth) is stopped when the loss function reaches a plateau. We build 1000 independent trees. The loss function reaches a plateau within a three depth of 30 for all the cases. The average prediction of all these decision trees is considered as the output of the RF model.

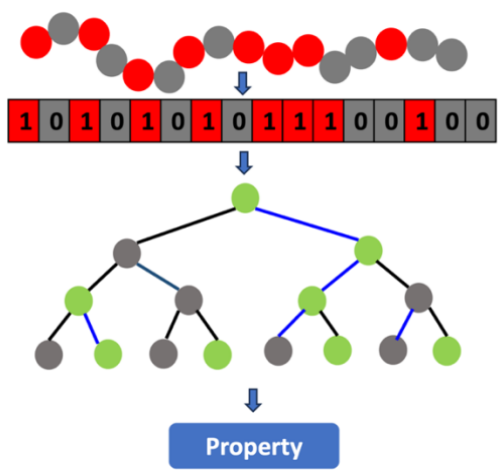

*Figure 2: A schematic representation of a tree model for predicting sequence-defined property of a copolymer. The copolymer is mapped to an array of binary number, which serve as the input to the tree. The average prediction from multiple such trees are considered as the output of a RF model.*

**XGBoost.** The XGBoost is a gradient boosted trees algorithm that provides a functional approximation of a correlation by optimizing specific loss functions as well as applying several regularization techniques.[31] These techniques penalize the model for being too complex, encouraging it to learn simpler and more generalizable patterns from the data. An XGBoost model is built in stages. It starts with a decision tree, and trains it on the data. Then, it analyzes the errors made by the first decision tree and creates a second decision tree focused on correcting the errors. This process continues sequentially, with each new tree learning from the error of the previous ones, as shown schematically in Figure

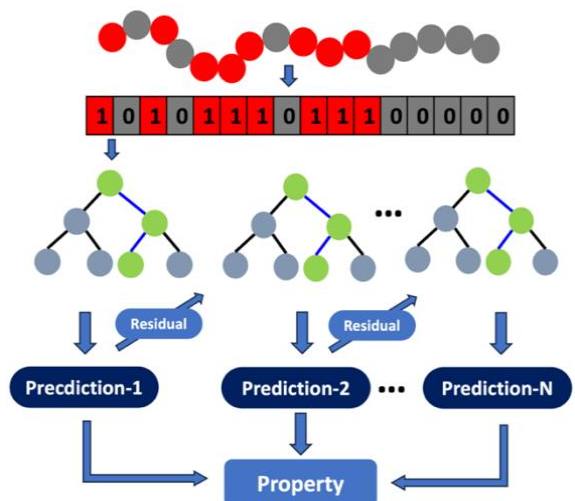

*Figure 3: A schematic representation of a XGBoost model for predicting the sequence-defined properties of a copolymer. The binary representations of copolymers and their AFEs are feeded to the first tree. The output of a tree serves as the input to the subsequent tree. The average prediction from all the trees is considered to be the output of the model.*

3. A total of 250 trees are connected in a sequence to build the model, beyond which the model's performance does not improve further for all the case studies. The depth of each tree is chosen to be 5.

**Supervised Deep Neural Network (DNN).** We build a fully connected feedforward backpropagation DNN for predicting sequence-defined properties of polymers, as schematically shown in Figure 4, similar to our recent work.[27] The DNN consists of two intermediate layers of perceptions, and these layers aid in polymer feature learning. The 1st and 2nd intermediate layers consist of 40 and 20 compute nodes, respectively. The number of intermediate layers and their size are determined based on multiple trials in order to reduce the loss function. The input layer has 20 nodes as the copolymer is represented by a binary number of lengths 20. The output layer of the DNN consists of one node that produces the AFE of the copolymer. All the nodes in the hidden layers and the output layer are activated using the rectified linear unit (ReLU)[32,33] activation function, which can be written as $f(x) = \max(0, x)$. This is a piecewise linear function that outputs the input directly if it is positive; otherwise, it will output zero. A node in the DNN receives the weighted sum of all the signals from all the nodes of its previous layer and outputs the activated signal, which is fed to all the nodes of its next layer. This feedforward process produces an output in the output layer node, which is the prediction of the model. The loss in the output layer node is backpropagated to estimate errors in all the compute nodes in the previous layers.[34] During the training, the weights are optimized using the Adam optimizer[35] to reduce the loss in all the compute nodes in the network. The training is stopped when the loss function reaches a plateau, which is about 150 cycles, beyond which the performance does not improve appreciably.

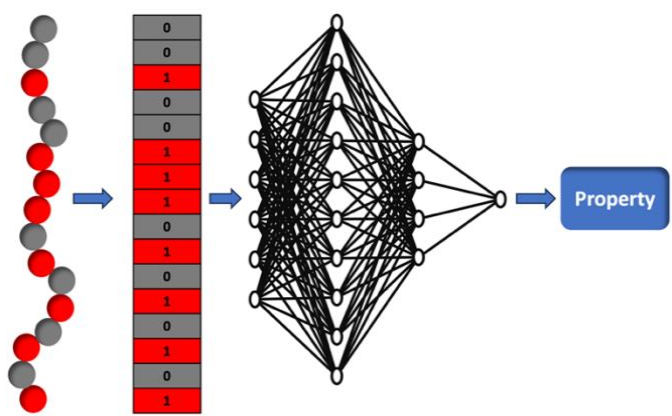

*Figure 4: A schematic representation of a DNN that predict sequence-defined property of a binary copolymer. The sequence of a copolymer is mapped to an array of binary number, which serve as the input to the model. There are two hidden layers in the DNN. A node in a hidden layer is connected to all the nodes of its preceding and succeeding layers. One output layer node that predict the adsorption free energy of the copolymer. The weights of all the connections are optimized during the training.*

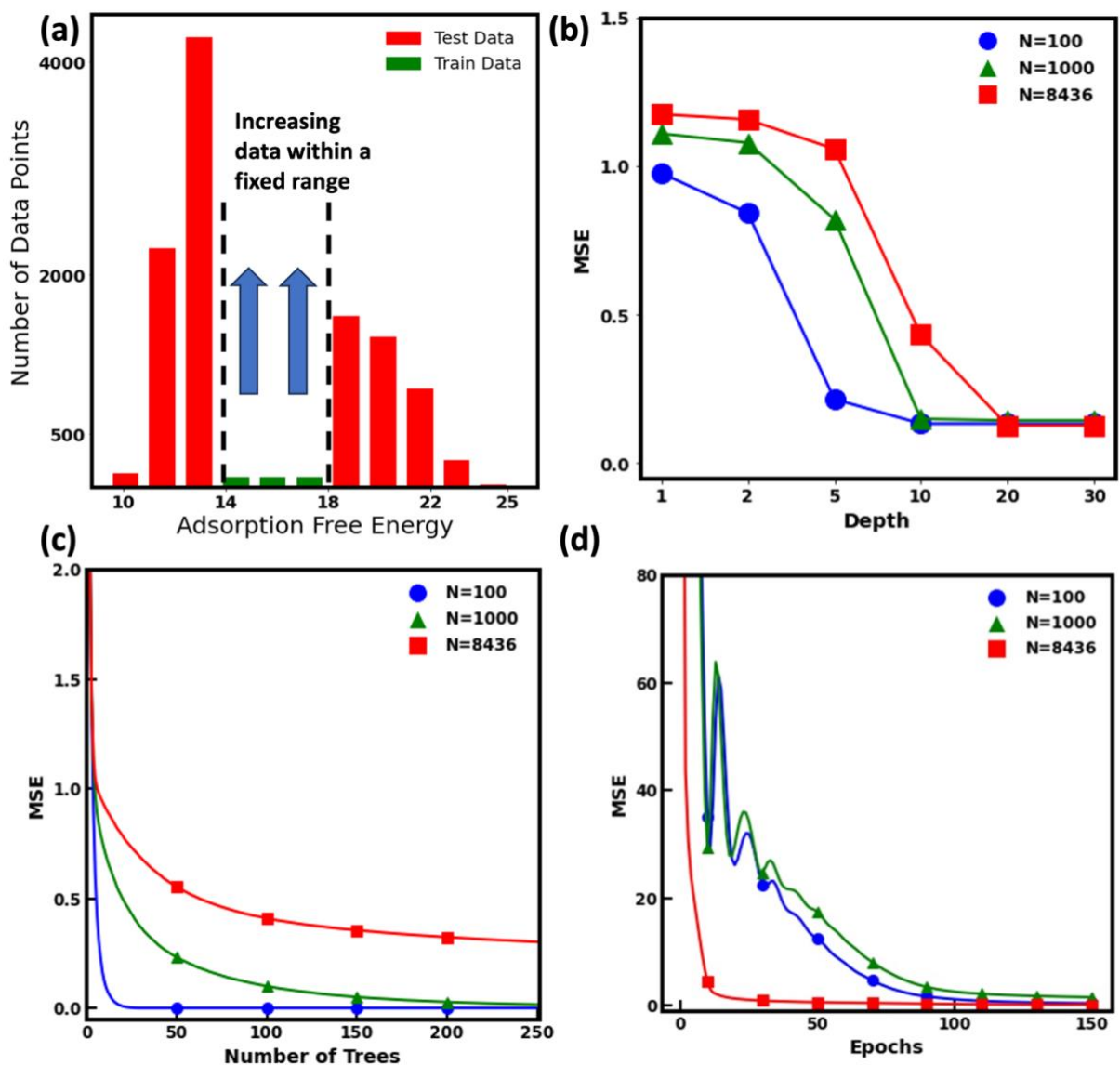

*Figure 5: Training ML models. The distribution of training and test data are shown in (a). The MSE is plotted as a function of the controlling variable in (b), (c), and (d), for RF, XG-boost and DNN, respectively for three representative cases.*

**Results and Discussion**

We curate two subsets of data for training and testing ML models. One is from the intermediate region of the AFE distribution as shown in Figure 5a for training the models. The data from the two extreme ends of the histogram are used for testing the model's extrapolation capability. The training and test sets consist of ~8000 and ~13000 data points, respectively. This partitioning ensures that we test a model's capability on truly outside the range of training data. We begin with a small amount of training data from the training set and systematically increase it. We evaluate the model' performance for the test data set for all the cases. Figure 5b-d show how different ML algorithms converge during their training for three different cases, each for a specific volume of training data. As the volume of training data increases, it requires deeper tree for achieving the convergence of a RF model. Similarly, more trees are required as the volume of training data increases for an XGBoost model to converge. Also, the MSE slightly increases as the number of training data increases for the XGBoot. The DNN learns faster with larger number of training data as the loss function reaches a plateau within 30 epochs for N=8436. The parity plots for two extreme cases for all three models – RF, XGBoost and DNN are shown in Figure 6. As shown in Figures 6 top panels, all three models perform poorly on the test data set when trained on 200 data points. As the training data increases, the performance of XGBoost and DNN on the test data set improves. Interestingly, the RF model's performance does not improve significantly despite of providing ~8000 data points for training. All three models perform well for the data points that are in the interpolative region. We note that the XGBoost and DNN models perform reasonably well for a few nearby points of the training data range for the case of 200 training data. But, they clearly fail in generalization for such lower number of training points. Their generalization capability improves upon adding more data as shown in the bottom panels of Figure 6.

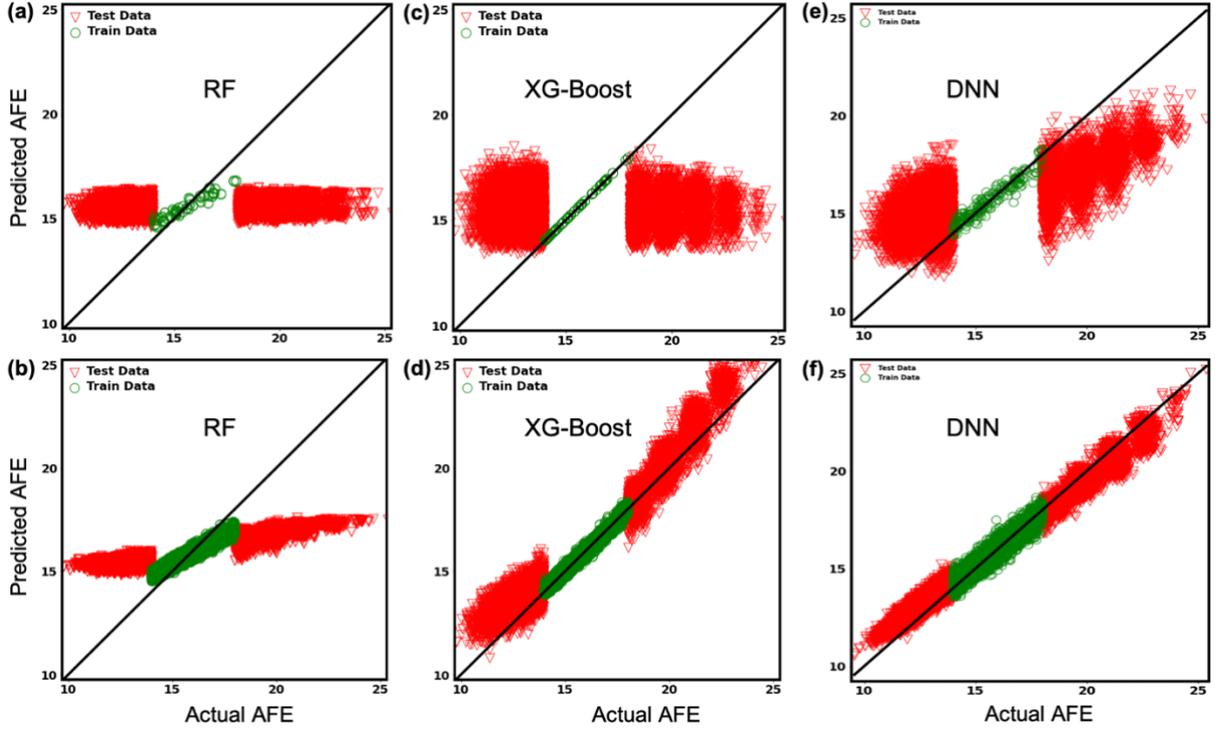

*Figure 6: Performance of ML models as the training data volume increase. The RF, XG-boost, and DNN models' parity plots are shown in (a,b), (c,d) and (e,f), respectively. The top panels correspond to 200 training data points and the bottom panels correspond to 8436 training data points.*

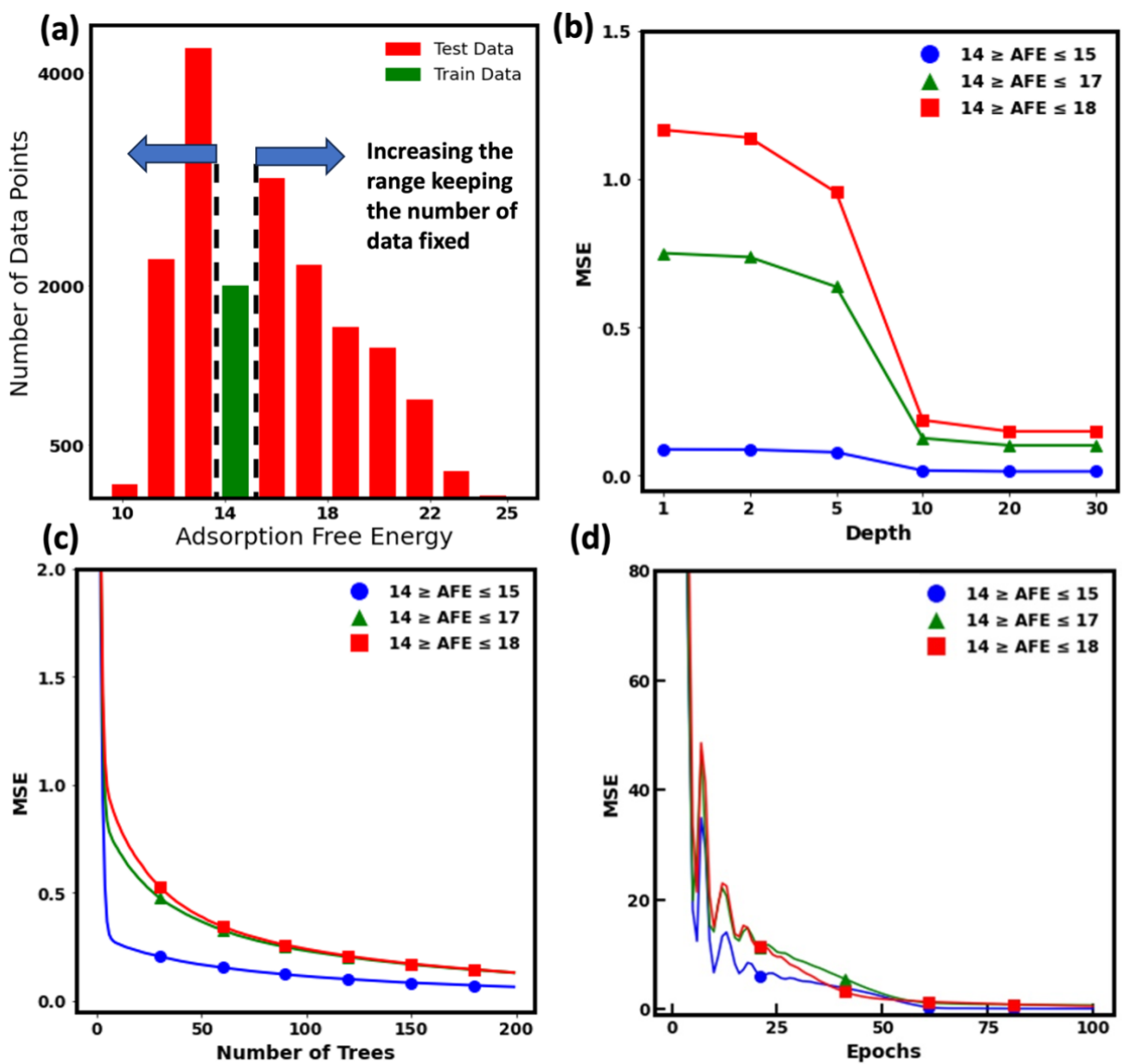

*Figure 7: Training ML models with varying range of data. The distribution of training and test data are shown in (a). The loss function – MSE is plotted as a function of controlling variable for the development of RF, XG-boost and DNN models in (b), (c), and (d), for respectively.*

Next, we build ML models wherein the range of the training data is systematically expanded as shown in Figure 7a. We test ML models' performance for increasing the range of training data. We fix the number of data points for training to be 1000. We randomly select 1000 data points for a specific range of AFE and use them to train the model and test the model's performance for all the points that are outside the range of the training data. This process is continued for three ranges of the AFE. The convergency of models during the training is shown in Figure 7b-d for all three cases. Both the RF and XG-boost converge to a slightly higher loss for a wider range of training data than the cases for a narrow range of training data. However, the DNN converges to the same point of the loss function. The performance of the three models is shown in Figure 8 for two extreme cases. The performance of all three models are very poor when the training data are selected for a free energy range of $14 \leq AFE \leq 15$,. As the training range become $14 \leq AFE \leq 18$, the performance of the XG-boost and DNN improves significantly. The 1000 data that spread across a wider range of the property space aid in learning more generalizable correlation for DNN and XG-boost. The RF performance remains poor in spite of a wider range of training data.

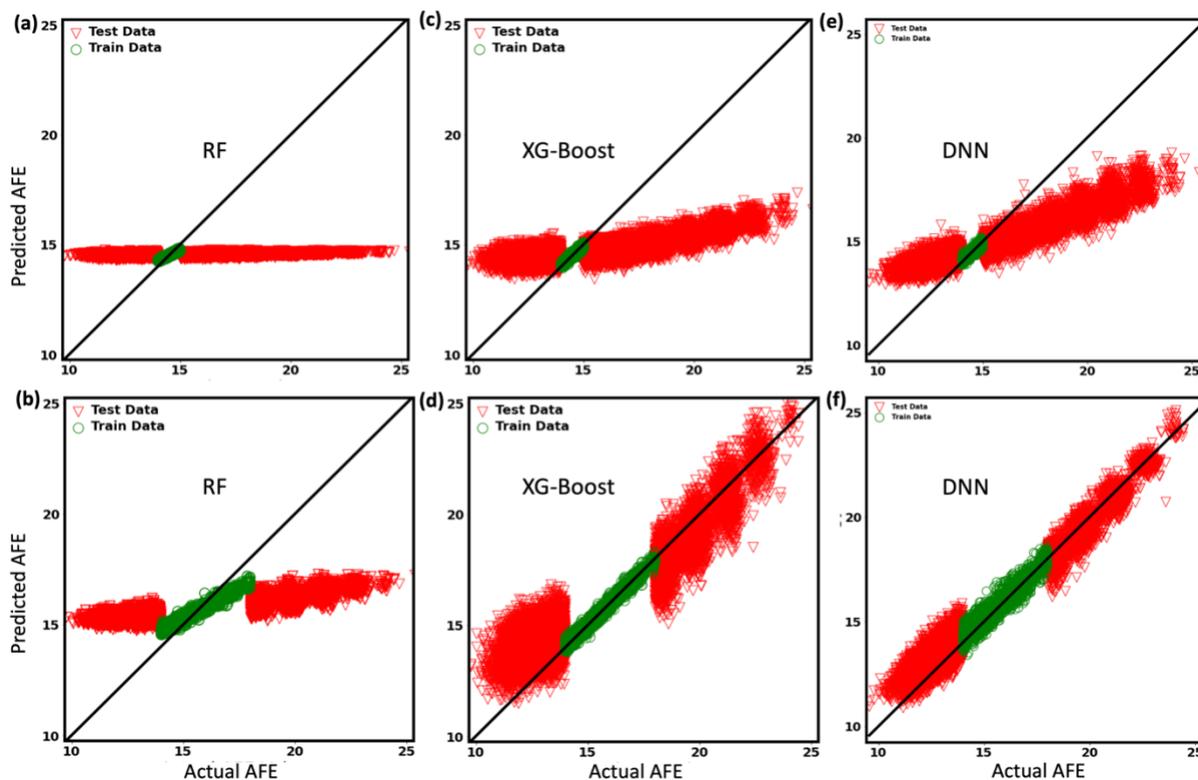

*Figure 8: ML models' performance for different training range of data. The number of training data is 1000 (green points). The top panel corresponds to a narrow range of training data ( $12 \leq AFE \leq 14$), and the bottom panel corresponds to a wider range of training data ($12 \leq AFE \leq 18$).*

The coefficient of determination for the above two extrapolation case studies are shown in Figure 9. For a given range, the extrapolation performance of DNN and XG-boost improves rapidly and reaches a plateau. Both the DNN and XG-boost reach their peak performance limit with about 2000 training data as shown in Figure 9a. The $R^2$ of the RF model for extrapolation is about 0.3 with 9000 training data. In case of fixed data points, the $R^2$ of the model monotonically increases as the range increases. The DNN shows an $R^2$ of 0.98 with the training data covering 20% of the range of the data set. For both the cases, the DNN shows superior performance over the XG-boost. By construct, a RF model learns similarity in structure of all the polymers. The output of a RF is the average property of those training data whose structures are similar to the test polymer. Whereas, the XG-boost and DNN learn correlations between dependent and independent variables of a data set. Now, given enough data or diversity of data, it seems possible to learn universal correlations using a DNN or XGBoost framework.

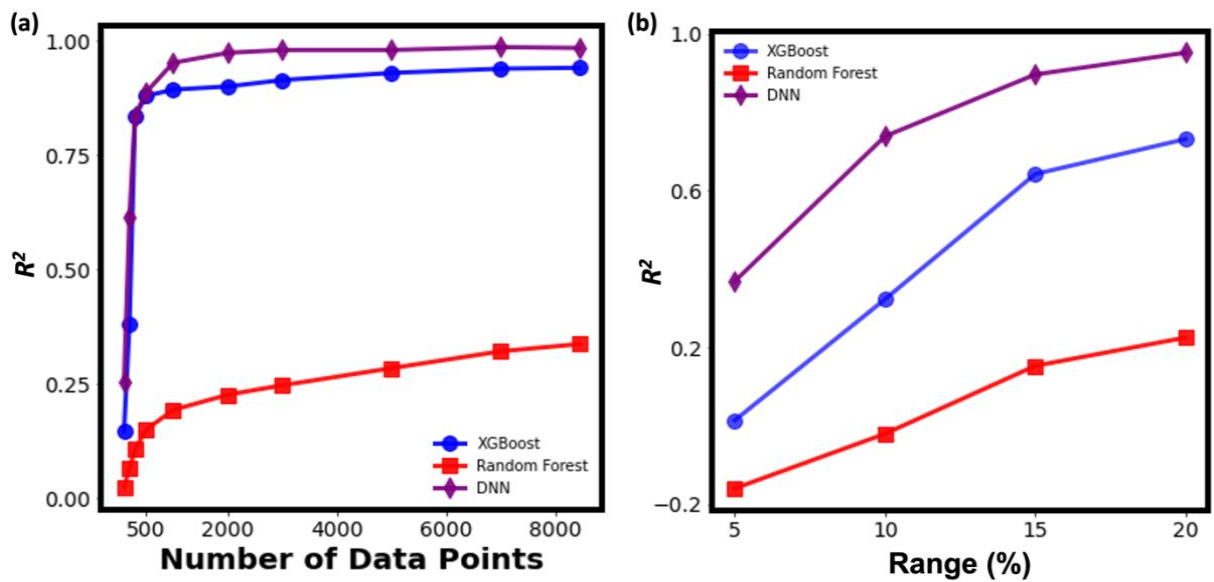

*Figure 9: The performance of ML models for extrapolation tasks. The coefficient of determination ($R^2$) is shown in (a) as a function of training data points that belong to a specific range of property. The $R^2$ is plotted as a function of the range of property for a fixed number of training data points.*

## IV. Conclusions

Machine learning and artificial intelligence bring huge promises to solve complex material science and engineering problems. The success of ML and AI-driven materials research relies on a deeper understanding of these tools, their generalizability, and their explainability. Here, we present a critical analysis on how ML models are generalizable in the context of polymer property predictions. We opt three ML algorithms – RF, XGboost and DNN and train them with homogeneous sequence-property data of specific distributions. We carefully analyze how they learn as the volume and range of data increases. We compute the transferability of these ML models across the sequence space of a binary copolymer, mainly when there is a training-target mismatch. We show that the XG-Boost and DNN can improve their extrapolation capability with an increasing amount of training data for a given range. Also, a smaller amount of training data collected over a wider range of properties leads to a higher fidelity extrapolation model. However, RF models are found to be unreliable for extrapolation tasks. Our analysis suggests that the scope of an RF is restricted to interpolations. Although we focus on sequence-property polymer models, we expect that these findings will be generalizable for other material systems. The work will have important implications in transferable machine learning model development for polymers and other materials.

**Data Availability**

The data that support the findings of this study are available from the corresponding author upon reasonable request.

**Notes**

The authors declare no competing financial interest.

**Acknowledgments**

The work is made possible by financial support from SERB, DST, Gov of India through a core research grant (CRG/2022/006926) and the National Supercomputing Mission's research grant (DST/NSM/R&D_HPC_Applications/2021/40). This research used the computational facility of the Center for Nanoscience Materials. Use of the Center for Nanoscale Materials, an Office of Science user facility, was supported by the U.S. Department of Energy, Office of Science, Office of Basic Energy Sciences, under Contract No. DE-AC02-06CH11357.

**References.**

(1) Patra, T. K. Data-Driven Methods for Accelerating Polymer Design. *ACS Polym. Au* **2022**, *2* (1), 8–26. https://doi.org/10.1021/acspolymersau.1c00035.

(2) Jackson, N. E.; Webb, M. A.; de Pablo, J. J. Recent Advances in Machine Learning towards Multiscale Soft Materials Design. *Curr. Opin. Chem. Eng.* **2019**, *23*, 106–114. https://doi.org/10.1016/j.coche.2019.03.005.

(3) Mannodi-Kanakkithodi, A.; Chandrasekaran, A.; Kim, C.; Huan, T. D.; Pilania, G.; Botu, V.; Ramprasad, R. Scoping the Polymer Genome: A Roadmap for Rational Polymer Dielectrics Design and Beyond. *Mater. Today* **2018**, *21* (7), 785–796. https://doi.org/10.1016/j.mattod.2017.11.021.

(4) Audus, D. J.; de Pablo, J. J. Polymer Informatics: Opportunities and Challenges. *ACS Macro Lett.* **2017**, *6* (10), 1078–1082. https://doi.org/10.1021/acsmacrolett.7b00228.

(5) Sattari, K.; Xie, Y.; Lin, J. Data-Driven Algorithms for Inverse Design of Polymers. *Soft Matter* **2021**. https://doi.org/10.1039/D1SM00725D.

(6) Bertinetto, C.; Duce, C.; Micheli, A.; Solaro, R.; Starita, A.; Tiné, M. R. Prediction of the Glass Transition Temperature of (Meth)Acrylic Polymers Containing Phenyl Groups by Recursive Neural Network. *Polymer* **2007**, *48* (24), 7121–7129. https://doi.org/10.1016/j.polymer.2007.09.043.

(7) Huan, T. D.; Mannodi-Kanakkithodi, A.; Ramprasad, R. Accelerated Materials Property Predictions and Design Using Motif-Based Fingerprints. *Phys. Rev. B* **2015**, *92* (1), 014106. https://doi.org/10.1103/PhysRevB.92.014106.

(8) Webb, M. A.; Jackson, N. E.; Gil, P. S.; Pablo, J. J. de. Targeted Sequence Design within the Coarse-Grained Polymer Genome. *Sci. Adv.* **2020**, *6* (43), eabc6216. https://doi.org/10.1126/sciadv.abc6216.

(9) Miccio, L. A.; Schwartz, G. A. From Chemical Structure to Quantitative Polymer Properties Prediction through Convolutional Neural Networks. *Polymer* **2020**, *193*, 122341. https://doi.org/10.1016/j.polymer.2020.122341.

(10) Drayer, W. F.; Simmons, D. S. Sequence Effects on the Glass Transition of a Model Copolymer System. *Macromolecules* **2022**. https://doi.org/10.1021/acs.macromol.2c00664.

(11) Tulsi, D. K.; Simmons, D. S. Hierarchical Shape-Specified Model Polymer Nanoparticles via Copolymer Sequence Control. *Macromolecules* **2022**, *55* (6), 1957–1969. https://doi.org/10.1021/acs.macromol.1c02215.

(12) Bale, A. A.; Gautham, S. M. B.; Patra, T. K. Sequence-Defined Pareto Frontier of a Copolymer Structure. *J. Polym. Sci. n/a* (n/a). https://doi.org/10.1002/pol.20220088.

(13) Meenakshisundaram, V.; Hung, J.-H.; Patra, T. K.; Simmons, D. S. Designing Sequence-Specific Copolymer Compatibilizers Using a Molecular-Dynamics-Simulation-Based Genetic Algorithm. *Macromolecules* **2017**, *50* (3), 1155–1166. https://doi.org/10.1021/acs.macromol.6b01747.

(14) Patra, T. K.; Loeffler, T. D.; Sankaranarayanan, S. K. R. Accelerating Copolymer Inverse Design Using Monte Carlo Tree Search. *Nanoscale* **2020**. https://doi.org/10.1039/D0NR06091G.

(15) Zhou, T.; Wu, Z.; Chilukoti, H. K.; Müller-Plathe, F. Sequence-Engineering Polyethylene–Polypropylene Copolymers with High Thermal Conductivity Using a Molecular-Dynamics-Based Genetic Algorithm. *J. Chem. Theory Comput.* **2021**. https://doi.org/10.1021/acs.jctc.1c00134.

(16) Chang, L.-W.; Lytle, T. K.; Radhakrishna, M.; Madinya, J. J.; Vélez, J.; Sing, C. E.; Perry, S. L. Sequence and Entropy-Based Control of Complex Coacervates. *Nat. Commun.* **2017**, *8* (1), 1273. https://doi.org/10.1038/s41467-017-01249-1.

(17) Meenakshisundaram, V.; Hung, J.-H.; Patra, T. K.; Simmons, D. S. Designing Sequence-Specific Copolymer Compatibilizers Using a Molecular-Dynamics-Simulation-Based Genetic Algorithm. *Macromolecules* **2017**, *50* (3), 1155–1166. https://doi.org/10.1021/acs.macromol.6b01747.

(18) Statt, A.; Casademunt, H.; Brangwynne, C. P.; Panagiotopoulos, A. Z. Model for Disordered Proteins with Strongly Sequence-Dependent Liquid Phase Behavior. *J. Chem. Phys.* **2020**, *152* (7), 075101. https://doi.org/10.1063/1.5141095.

(19) Statt, A.; Kleeblatt, D. C.; Reinhart, W. F. Unsupervised Learning of Sequence-Specific Aggregation Behavior for a Model Copolymer. *Soft Matter* **2021**, *17* (33), 7697–7707. https://doi.org/10.1039/D1SM01012C.

(20) Bhattacharya, D.; Kleeblatt, D. C.; Statt, A.; Reinhart, W. F. Predicting Aggregate Morphology of Sequence-Defined Macromolecules with Recurrent Neural Networks. *Soft Matter* **2022**, *18* (27), 5037–5051. https://doi.org/10.1039/D2SM00452F.

(21) Patel, R. A.; Borca, C. H.; Webb, M. A. Featurization Strategies for Polymer Sequence or Composition Design by Machine Learning. *Mol. Syst. Des. Eng.* **2022**, *7* (6), 661–676. https://doi.org/10.1039/D1ME00160D.

(22) Tao, L.; Byrnes, J.; Varshney, V.; Li, Y. Machine Learning Strategies for the Structure-Property Relationship of Copolymers. *iScience* **2022**, *25* (7), 104585. https://doi.org/10.1016/j.isci.2022.104585.

(23) Shi, J.; Quevillon, M. J.; Amorim Valença, P. H.; Whitmer, J. K. Predicting Adhesive Free Energies of Polymer–Surface Interactions with Machine Learning. *ACS Appl. Mater. Interfaces* **2022**, *14* (32), 37161–37169. https://doi.org/10.1021/acsami.2c08891.

(24) Kremer, K.; Grest, G. S. Dynamics of Entangled Linear Polymer Melts: A Molecular-dynamics Simulation. *J. Chem. Phys.* **1990**, *92* (8), 5057–5086. https://doi.org/10.1063/1.458541.

(25) Grest, G. S.; Kremer, K. Molecular Dynamics Simulation for Polymers in the Presence of a Heat Bath. *Phys. Rev. A* **1986**, *33* (5), 3628–3631. https://doi.org/10.1103/PhysRevA.33.3628.

(26) Everaers, R.; Karimi-Varzaneh, H. A.; Fleck, F.; Hojdis, N.; Svaneborg, C. Kremer–Grest Models for Commodity Polymer Melts: Linking Theory, Experiment, and Simulation at the Kuhn Scale. *Macromolecules* **2020**, *53* (6), 1901–1916. https://doi.org/10.1021/acs.macromol.9b02428.

(27) Himanshu; Chakraborty, K.; Patra, T. K. Developing Efficient Deep Learning Model for Predicting Copolymer Properties. *Phys. Chem. Chem. Phys.* **2023**, *25* (37), 25166–25176. https://doi.org/10.1039/D3CP03100D.

(28) McInnes, L.; Healy, J.; Melville, J. UMAP: Uniform Manifold Approximation and Projection for Dimension Reduction. arXiv September 17, 2020. https://doi.org/10.48550/arXiv.1802.03426.

(29) *Keras: the Python deep learning API*. https://keras.io/ (accessed 2020-10-05).

(30) Breiman, L. Random Forests. *Mach. Learn.* **2001**, *45* (1), 5–32. https://doi.org/10.1023/A:1010933404324.

(31) Chen, T.; Guestrin, C. XGBoost: A Scalable Tree Boosting System. In *Proceedings of the 22nd ACM SIGKDD International Conference on Knowledge Discovery and Data Mining*; KDD '16; Association for Computing Machinery: New York, NY, USA, 2016; pp 785–794. https://doi.org/10.1145/2939672.2939785.

(32) Glorot, X.; Bordes, A.; Bengio, Y. Deep Sparse Rectifier Neural Networks. In *Proceedings of the Fourteenth International Conference on Artificial Intelligence and Statistics*; JMLR Workshop and Conference Proceedings, 2011; pp 315–323.

(33) Nair, V.; Hinton, G. E. Rectified Linear Units Improve Restricted Boltzmann Machines. In *Proceedings of the 27th International Conference on International Conference on Machine Learning*; ICML'10; Omnipress: Madison, WI, USA, 2010; pp 807–814.

(34) LeCun, Y.; Bottou, L.; Orr, G. B.; Müller, K.-R. Efficient BackProp. In *Neural Networks: Tricks of the Trade*; Orr, G. B., Müller, K.-R., Eds.; Lecture Notes in Computer Science; Springer Berlin Heidelberg, 1998; pp 9–50. https://doi.org/10.1007/3-540-49430-8_2.

(35) Kingma, D. P.; Ba, J. Adam: A Method for Stochastic Optimization. arXiv January 29, 2017. https://doi.org/10.48550/arXiv.1412.6980.